\documentclass[11pt,showpacs,preprintnumbers,nofootinbib]{revtex4} 

\usepackage{graphicx}
\usepackage{dcolumn}
\usepackage{bm}
\usepackage{amssymb}
\usepackage{amsmath}
\usepackage{epsfig}    
\usepackage{color}    

\def\beq{\begin{equation}}
\def\eeq{\end{equation}}
\def\bea{\begin{array}}
\def\eea{\end{array}}
\def\be{\begin{equation}}
\def\ee{\end{equation}}
\def\ba{\begin{eqnarray}}
\def\ea{\end{eqnarray}}

\def\to{\rightarrow}

\def\[{\left[}
\def\]{\right]}
\def\({\left(}
\def\){\right)}

\def\sm0{{\widetilde{m}_0}}

\def\U1em{{U(1)_{\rm em}}}
\def\to{\rightarrow}

\def\sq2{\sqrt{2}}

\def\ee{e^+e^-}

\def\End{\end{document}}

\def\fsl#1{\setbox0=\hbox{$#1$}                 
   \dimen0=\wd0                                 
   \setbox1=\hbox{/} \dimen1=\wd1               
   \ifdim\dimen0>\dimen1                        
      \rlap{\hbox to \dimen0{\hfil/\hfil}}      
      #1                                        
   \else                                        
      \rlap{\hbox to \dimen1{\hfil$#1$\hfil}}   
      /                                         
   \fi}

\begin{document} 

\title{Renormalization of the Higgs sector in the triplet model}
\preprint{KANAZAWA-12-04, UT-HET 067}
\author{%
{\sc Mayumi Aoki\,$^1$, Shinya Kanemura\,$^2$, Mariko Kikuchi\,$^2$,   
    and Kei Yagyu\,$^{2}$\footnote{Adrress after April 2012, National Central University, Taiwan.}}}
\affiliation{
$^1$Institute~for~Theoretical~Physics,~Kanazawa~University,~Kanazawa~920-1192,~Japan\\
$^2$Department of Physics, University of Toyama, \\3190 Gofuku, Toyama 930-8555, Japan\\}
\begin{abstract}
We study radiative corrections to the mass spectrum and the triple Higgs boson coupling 
in the model with an additional $Y=1$ triplet field. 
In this model, the vacuum expectation value for the triplet field is strongly constrained 
from the electroweak precision data,
under which 
characteristic mass spectrum appear at the tree level; i.e., 
$m_{H^{++}}^2-m_{H^+}^2\simeq m_{H^+}^2-m_A^2$ and $m_A^2\simeq m_H^2$, 
where 
the CP-even ($H$), the CP-odd ($A$) and  
the doubly-charged ($H^{\pm\pm}$) as well as the singly-charged ($H^\pm$) Higgs bosons are 
the triplet-like. 
We evaluate how the tree-level formulae are modified at the one-loop level. 
The $hhh$ coupling for the standard model-like Higgs boson ($h$) is also 
calculated at the one-loop level. 
One-loop corrections to these quantities can be large enough 
for identification of the model by future precision data 
at the LHC or the International Linear Collider.

\end{abstract}

\maketitle


\section{Introduction}
The Higgs boson search is underway at the CERN LHC. 
By the recent results, 
the allowed regions of the Higgs boson mass has been constrained to be 
117.5-118.5 GeV, 122.5-129 GeV and larger than 539 GeV at 95\% Confidence Level (CL) 
at the ATLAS~\cite{ATLAS}, 
and 114.4-127.5 GeV and larger than 600 GeV at 95\% CL at the CMS~\cite{CMS}. 
Since the electroweak precision data at the LEP suggest that 
the Higgs boson mass is less than around 160 GeV at 95\% CL~\cite{LEP} assuming the Standard Model (SM), 
we expect that a light (SM-like) Higgs boson will be discovered with the mass of around 125 GeV soon.

On the other hand, it is well known that several phenomena such as 
tiny neutrino masses~\cite{neutrino-oscillation}, existence of dark matter~\cite{DM} 
and baryon asymmetry of the Universe~\cite{BAU} cannot be explained in the SM. 
Because there is no strong motivation to the SM (minimal) Higgs sector, 
we may consider extensions of the Higgs sector in order to explain these phenomena. 
First, by imposing an unbroken discrete symmetry to the Higgs sector, we can obtain dark matter candidates, such as
the inert doublet model~\cite{inert_doublet}. 
Second, the type-II seesaw model~\cite{typeII}, where a Higgs triplet field is added to the SM, can generate 
neutrino masses at the tree level.  
Radiative seesaw models~\cite{zee,zee-2loop,krauss,ma,aks} can also explain neutrino masses at loop levels, 
where additional scalar bosons, e.g., charged scalar bosons, are running in the loop. 
Finally, the scenario based on the electroweak baryogenesis
can explain the baryon asymmetry of the Universe 
by an additional CP-phase and nondecoupling effects in extended Higgs sectors~\cite{ewbg-thdm,non_dec} . 

Discoveries of additional scalar bosons such as charged Higgs bosons 
will be direct evidence of extended Higgs sectors. 
In addition, as indirect signatures,  
detecting the deviation of the Higgs boson couplings with 
gauge bosons $hVV$, those with quarks and leptons $h\bar{f}f$ and Higgs selfcoupling constant $hhh$, 
where $h$ is the SM-like Higgs boson, 
from those predicted values in the SM
can also be evidence of extended Higgs sectors. 
In particular, once $h$ is discovered at the LHC in near future, 
precision measurements of these coupling constants at the LHC and at the International Linear Collider (ILC)
turns to be very important. 
Therefore, in order to distinguish various extended Higgs sectors, 
it is necessary to 
prepare precise calculations for the Higgs boson interactions  
including radiative corrections.

In this Letter, we construct the renormalization scheme for the one-loop calculation of the observables in the Higgs sector 
such as mass spectrum and coupling constants 
in the Higgs triplet model (HTM), where the 
hypercharge $Y=1$ Higgs triplet field is added to the SM. 
The HTM is motivated to generate tiny neutrino masses by the type-II seesaw mechanism. 
This model is unique because the electroweak rho parameter deviates from unity at the tree level. 
In such a model, unlike the SM, four independently measured parameters such as 
$\alpha_{\text{em}}$, $G_F$, $m_Z$ and $\sin^2\theta_W$ 
are necessary to describe the electroweak observables~\cite{blank_hollik}.  
In the model with the $Y=0$ Higgs triplet field, 
the rho parameter and the W boson mass were calculated at the one-loop level in the on-shell scheme
in Ref.~\cite{blank_hollik,real_triplet}. 
Recently, in the HTM, 
one-loop corrections to the rho parameter as well as the W boson mass have been studied~\cite{ky}, 
and it is clarified that relatively large mass differences among the triplet-like Higgs bosons 
are favored by the precision data at the LEP/SLC~\cite{LEP,SLC}. 
However, radiative corrections to the Higgs sector with renormalization have not been studied in the HTM so far.  

In the HTM, there are seven physics scalar states; i.e., 
the doubly-charged $H^{\pm\pm}$, 
the singly-charged $H^\pm$, a CP-odd $A$ as well as two CP-even ($h$ and $H$) scalar states. 
When the vacuum expectation value (VEV) of the triplet field $v_\Delta$ is much smaller 
than that of the doublet field $v_\phi$ as required by the electroweak precision data, $h$ behaves as the SM-like Higgs boson while 
$H^{\pm\pm}$, $H^\pm$, $H$ and $A$ 
are the triplet-like ones. 
Phenomenology for the HTM 
has been studied in Refs.~\cite{HTM_pheno_wo1,HTM_pheno_wo2,HTM_pheno_nmass,HTM_pheno_w1,HTM_pheno_w2,aky_triplet}.  
When the triplet-like Higgs bosons are degenerated in mass, 
$H^{++}$ mainly decays into the same sign dilepton or the diboson depending on the 
magnitude of $v_\Delta$. 
On the other hand, in the case with the mass difference among the triplet-like fields, 
cascade decays of the triplet-like Higgs bosons can be dominant~\cite{HTM_pheno_w1,HTM_pheno_w2,aky_triplet}. 
For example, when $H^{++}$ ($H$ or $A$) is the heaviest of all the triplet-like Higgs bosons, 
the main decay mode of $H^{++}$ ($H$ or $A$) can be $H^{++}\to H^+W^{+(*)}\to H W^{+(*)}W^{+(*)}$ or $A W^{+(*)}W^{+(*)}$
($H$ or $A\to H^\pm W^{\mp(*)}\to H^{\pm\pm} W^{\mp(*)}W^{\mp(*)}$).   
In any case, measuring the masses of the triplet-like Higgs bosons  
is important to test the HTM. 

For $v_\Delta^2/v_\phi^2\ll 1$, characteristic relationships among the masses of the triplet-like Higgs bosons 
are predicted at the tree level~\cite{aky_triplet}:
\begin{align}
&m_{H^{++}}^2-m_{H^+}^2= m_{H^+}^2-m_A^2,\label{eq:intro1}\\
&m_A^2= m_H^2, \label{eq:intro2}
\end{align}
up to $v_\Delta^2/v_\phi^2$, where $m_{H^{++}}$, $m_{H^+}$, $m_H$ and $m_A$ are the masses of $H^{\pm\pm}$, 
$H^\pm$, $H$ and $A$, respectively. 
These formulae are useful to distinguish the model from the others which also contain charged scalar bosons 
when the masses of the triplet-like Higgs bosons are measured with sufficient accuracy. 
In particular, in order to compare the mass spectrum with future precision measurements, 
it is very important to evaluate the radiative correction to the above tree-level formulae.  
We first define $\Delta R\equiv (m_{H^{++}}^2-m_{H^+}^2)/(m_{H^+}^2-m_A^2)-1$ in order to
investigate how the mass formulae in Eq.~({\ref{eq:intro1}}) are modified by the quantum effect. 
We find that the magnitude for $\Delta R$ can be as large as $10\%$ when we take parameter sets favored by the electroweak precision data: 
$v_\Delta^2/v_\phi^2\ll 1$, $m_{H^{++}}= \mathcal{O}(100)$ GeV and $m_{H^+}-m_{H^{++}}\simeq 100$ GeV~\cite{ky}. 

The theoretical prediction on the triple Higgs boson coupling $hhh$  
is also quite important because
the Higgs potential is reconstructed by measuring the mass of $h$ and the coupling of $hhh$. 
In the case of the two Higgs doublet model,  
radiative corrections to the coupling of $hhh$ have been studied by using 
the effective potential method in Ref.~\cite{KKOSY} 
as well as the diagrammatical method in the on-shell scheme in Ref.~\cite{KOSY}, in which 
the renormalized $hhh$ coupling 
can deviate from the SM prediction by about 100\% due to the nondecoupling property of 
the additional scalar bosons without contradiction of the perturbative unitarity~\cite{PU_thdm}. 
We here calculate the renormalized $hhh$ coupling in the HTM, and find that the deviation of this coupling 
from the SM prediction can also be of $O(100)\%$ in the parameter choice of 
$v_\Delta^2/v_\phi^2\ll 1$, $m_{H^{++}}= \mathcal{O}(100)$ GeV and $m_{H^+}-m_{H^{++}}\simeq 100$ GeV.

\section{Model}
The HTM is one of the minimal extension of the SM, where 
the Higgs triplet field $\Delta$ with $Y=1$ is added to the SM. 
The most general  
Higgs potential under the $SU(2)_L\times U(1)_Y$ gauge invariance 
is given by 
\begin{eqnarray}
 V &=& m^2 \Phi^\dagger \Phi + M^2 {\rm Tr}(\Delta^\dagger \Delta) +
  [ \mu \Phi^T i \tau_2 \Delta^\dagger \Phi + {\rm h.c.} ] \nonumber \\
&& + \lambda_1 (\Phi^\dagger \Phi)^2 + \lambda_2 \left[{\rm
                                                  Tr}(\Delta^\dagger
                                                  \Delta) \right]^2
  + \lambda_3 {\rm Tr}\left[(\Delta^\dagger \Delta)^2\right]
  +\lambda_4 (\Phi^\dagger\Phi) {\rm Tr}(\Delta^\dagger \Delta)
  +\lambda_5 \Phi^\dagger \Delta \Delta^\dagger \Phi. \label{eq:pot}
 \end{eqnarray}  
The Higgs fields are parameterized as
\begin{eqnarray}
    \Phi = \left[\begin{array}{c}
            w^+ \\
       \frac{1}{\sqrt{2}}(\varphi^0+v_\phi + i z^0)
    \end{array}
           \right],\quad
   \Delta = \left[\begin{array}{cc}
 \frac{\Delta^+}{\sqrt{2}} & \Delta^{++} \\
 \Delta^0 & - \frac{\Delta^+}{\sqrt{2}}  \\
                  \end{array}\right],\quad 
\text{with }\Delta^0 = \frac{1}{\sqrt{2}} (\delta^0+v_\Delta + i \xi^0), 
\end{eqnarray}
where $v_\phi$ and $v_\Delta$ are the VEVs of the doublet field and the triplet field, respectively, which  
satisfy $v^2\equiv v_\phi^2+2v_\Delta^2=(\sqrt{2} G_F)^{-1}\simeq (246\text{ GeV})^2$. 
When the triplet field $\Delta$ carries lepton number of two units ($L=-2$),  
the $\mu$ term breaks the lepton number, so that  
it is the source of Majorana masses of neutrinos. 
 
The tree-level contributions to the tadpoles for the component fields $\varphi^0$ and $\delta^0$ are 
given by
\begin{eqnarray}
 T_\phi &=& - m^2 v_\phi + \sqrt{2} \mu v_\Delta v_\phi
-\lambda_1 v_\phi^3  -\frac{1}{2}(\lambda_4+\lambda_5)v_\Delta^2 v_\phi, \\
 T_\Delta &=& -M^2 v_\Delta + \frac{\mu v_\phi^2}{\sqrt{2}} 
-(\lambda_2+\lambda_3) v_\Delta^3 - \frac{1}{2}(\lambda_4+\lambda_5)
v_\phi^2 v_\Delta.
\end{eqnarray}
By using the stationary conditions $T_\phi=T_\Delta=0$, we obtain 
\begin{eqnarray}
 m^2 &=& \sqrt{2} \mu v_\Delta - \lambda_1 v_\phi^2
  -\frac{1}{2} (\lambda_4+\lambda_5) v_\Delta^2 , \label{eq:vc1}\\
 M^2 &=&  M_\Delta^2 -
  (\lambda_2+\lambda_3) v_\Delta^2
  - \frac{1}{2}(\lambda_4+\lambda_5)v_\phi^2,   \label{eq:vc2}
 \end{eqnarray}
where $M_\Delta^2 = \frac{\mu v_\phi^2}{\sqrt{2} v_\Delta}$. 
The mass of the doubly-charged scalar bosons $\Delta^{\pm\pm}$ $(\equiv H^{\pm\pm})$ 
is given as
\begin{eqnarray}
 m_{H^{++}}^2 = M_\Delta^2 - \lambda_3 v_\Delta^2 - \frac{1}{2}\lambda_5
  v_\phi^2. \label{eq:mHpp}
\end{eqnarray}
The mass matrices can be expressed 
for the singly-charged scalar states in the basis of $(w^+,\Delta^+)$ 
and the CP-odd scalar states in the basis of $(z^0, \xi^0)$ as
 \begin{align}
  M_{\rm charged}^2 = \left(M_\Delta^2 - \frac{1}{4}\lambda_5
                     v_\phi^2\right)
  \left[ \begin{array}{cc}
              \frac{2 v_\Delta^2}{v_\phi^2}  &
             -  \frac{\sqrt{2} v_\Delta}{v_\phi} \\ 
             -  \frac{\sqrt{2} v_\Delta}{v_\phi} &    1  \\
         \end{array}\right],\quad 
  M_{\rm CP-odd}^2 = M_\Delta^2 
  \left[ \begin{array}{cc}
              \frac{4 v_\Delta^2}{v_\phi^2}  &
             -  \frac{2 v_\Delta}{v_\phi} \\
             -  \frac{2 v_\Delta}{v_\phi} &    1  \\
         \end{array}\right].
  \end{align}
The mass eigenstates of the scalar fields are obtained by 
diagonalizing these mass matrices by the mixing angle $\beta$ and $\beta'$ as
\begin{align}
  \left[\begin{array}{c}
        w^\pm \\
        \Delta^\pm \\
        \end{array}\right]
  =
  \left[\begin{array}{cc}
        \cos\beta & - \sin\beta \\
        \sin\beta & \cos\beta   \\
        \end{array}\right]
  \left[\begin{array}{c}
        G^\pm \\
        H^\pm \\
        \end{array}\right], \quad
  \left[\begin{array}{c}
        z^0 \\
        \xi^0 \\
        \end{array}\right]
  =
  \left[\begin{array}{cc}
        \cos\beta' & - \sin\beta' \\
        \sin\beta' & \cos\beta'   \\
        \end{array}\right]
  \left[\begin{array}{c}
        G^0 \\
        A \\
        \end{array}\right], 
\end{align}
where $G^\pm$ and $G^0$ are the Nambu-Goldstone bosons which are absorbed by the longitudinal component of the weak gauge bosons, 
and the mixing angles are given by
\begin{eqnarray}
 \tan\beta= \frac{\sqrt{2} v_\Delta}{v_\phi},\quad   \tan\beta'=\frac{2 v_\Delta}{v_\phi}.
\end{eqnarray}
We then obtain the masses of the singly-charged scalar bosons and the CP-odd scalar boson as
\begin{eqnarray}
 m_{H^+}^2 &=& \left(M_\Delta^2 - \frac{1}{4} \lambda_5 v_\phi^2\right)
  \left(1 + \frac{2 v_\Delta^2}{v_\phi^2} \right),   \label{eq:mHp}\\
 m_{A}^2 &=& M_\Delta^2 \left(1 + \frac{4 v_\Delta^2}{v_\phi^2} \right).\label{eq:mA}
\end{eqnarray}
For the CP-even scalar states, the mass matrix is given in the basis of ($\varphi^0$, $\delta^0$) 
by 
\begin{align}
M_{\rm CP-even}^2=
\left[
\begin{array}{cc}
2v_\phi^2\lambda_1 & -\frac{2v_\Delta}{v_\phi}M_\Delta^2+v_\phi v_\Delta(\lambda_4+\lambda_5)\\
-\frac{2v_\Delta}{v_\phi}M_\Delta^2+v_\phi v_\Delta(\lambda_4+\lambda_5) &M_\Delta^2+2v_\Delta^2(\lambda_2+\lambda_3)
\end{array}\right]. \label{eq:meven}
\end{align}
By the diagonalization of this matrix with the mixing angle $\alpha$ as
\begin{align}
  \left[\begin{array}{c}
        \varphi^0 \\
        \delta^0 \\
        \end{array}\right]
  =
  \left[\begin{array}{cc}
        \cos\alpha & - \sin\alpha \\
        \sin\alpha & \cos\alpha   \\
        \end{array}\right]
  \left[\begin{array}{c}
        h \\
        H \\
        \end{array}\right], 
\end{align}
we obtain the masses of the CP-even Higgs bosons and the mixing angle: 
\begin{eqnarray}
 m_h^2 &=&   (M_{\rm CP-even}^2)_{11} \cos^2 \alpha+ (M_{\rm
  CP-even}^2)_{22} \sin^2\alpha + 2 (M_{\rm CP-even}^2)_{12}
  \cos\alpha\sin\alpha,
  \label{eq:mh}\\
 m_H^2 &=& (M_{\rm CP-even}^2)_{11} \sin^2 \alpha + (M_{\rm
  CP-even}^2)_{22} \cos^2\alpha - 2 (M_{\rm CP-even}^2)_{12}
  \cos\alpha\sin\alpha, \label{eq:mH}\\
 \tan 2\alpha &=& \frac{2 (M_{\rm CP-even}^2)_{12}}
  {(M_{\rm CP-even}^2)_{11}-(M_{\rm CP-even}^2)_{22}}, \label{eq:tan2a}
 \end{eqnarray}
where $(M_{\rm CP-even}^2)_{ij}$, $(i,j=1,2)$ are elements of the mass matrix in Eq.~(\ref{eq:meven}). 
From Eqs.~(\ref{eq:vc1}), (\ref{eq:vc2}), (\ref{eq:mHpp}), (\ref{eq:mHp}), (\ref{eq:mA}), (\ref{eq:mh}), (\ref{eq:mH}) and (\ref{eq:tan2a}), 
eight parameters: $m$, $M$, $\mu$ and $\lambda_1$-$\lambda_5$ in the Higgs potential can be translated into five mass parameters: 
$m_{H^{++}}$, $m_{H^{+}}$, $m_H$, $m_A$ and $m_{h}$, one mixing angle $\alpha$ and two VEVs: $v_\phi$ and $v_\Delta$. 
In the next section, we take $v$ as one of the input parameters instead of $v_\phi$. 

Masses of neutrinos are generated through the 
Yukawa interaction: 
\begin{align}
\mathcal{L}_{\nu}&=h_{ij}\overline{L_L^{ic}}i\tau_2\Delta L_L^j+\text{h.c.},\label{yn}
\end{align}
where $h_{ij}$ is a $3\times 3$ symmetric complex matrix,  
$L_L^i$ is the $i$-th generation of the left-handed lepton doublet. 
After the electroweak symmetry breaking, 
the neutrino masses are obtained as
\begin{align}
(\mathcal{M}_\nu)_{ij}=\sqrt{2}h_{ij}v_\Delta=h_{ij}\frac{\mu v_\phi^2}{M_\Delta^2}.  \label{eq:mn}
\end{align}
It can be seen that when the lepton number violating parameter $\mu$ is taken to be of $\mathcal{O}(0.1-1)$ eV with $h_{ij}=\mathcal{O}(1)$, 
we can take $M_\Delta$ to be $\mathcal{O}(100-1000)$ GeV with satisfying $(\mathcal{M}_\nu)_{ij}=\mathcal{O}(0.1)$ eV 
which is required by the data. 
In such a case, the HTM can be tested at TeV-scale collider experiments. 
The smallness of $\mu$ is just the assumption in the minimal version of the HTM. 
However, if the parameter $\mu$ is forbidden at the tree level but is generated at higher orders of perturbation, 
such a tiny value of $\mu$ may be explained in a natural way~\cite{kanemura_sugiyama}.

In the HTM, the electroweak rho parameter $\rho$ deviates from unity at the tree level: 
\begin{align}
\rho \equiv \frac{m_W^2}{m_Z^2\cos^2\theta_W}=\frac{1+\frac{2v_\Delta^2}{v_\phi^2}}{1+\frac{4v_\Delta^2}{v_\phi^2}}. 
\end{align}
The experimental value of the rho parameter $\rho_{\text{exp}}$ is quite close to unity: 
$\rho_{\text{exp}}=1.0008_{-0.0007}^{+0.0017}$~\cite{pdg}, 
so that the value of $v_\Delta$ is bounded from above: $v_\Delta\lesssim 8$ GeV. 
Therefore, $v_\Delta^2/v_\phi^2\lesssim 0.001$ is phenomenologically acceptable. 
In such a case, since $\beta$, $\beta'$ and $\alpha$ are near zero, 
the scalar bosons $H^{\pm\pm}$, $H^\pm$, $A$ and $H$ can be regarded as the triplet-like Higgs bosons 
while $h$ behaves the SM-like Higgs boson. 
Neglecting terms with $v_\Delta^2/v_\phi^2$, 
there appear characteristic mass formulae as 
\begin{align}
m_{H^{++}}^2-m_{H^{+}}^2 &= m_{H^+}^2-m_A^2 \left(= -\frac{\lambda_5}{4}v_\phi^2\right), \label{mass1}\\
m_H^2 &= m_A^2\left(=M_\Delta^2 \right). \label{mass2}
\end{align}
The relation in Eq.~(\ref{mass1}) can also be described as 
\begin{align}
R&\equiv \frac{m_{H^{++}}^2-m_{H^{+}}^2}{m_{H^+}^2-m_A^2}=1.  \label{R_def}
\end{align}

The mass formulae Eqs.~(\ref{mass1}) and (\ref{mass2}) can be understood 
by considering global symmetries in the Higgs potential in Eq.~(\ref{eq:pot}), which is 
invariant under the $SU(2)_L\times U(1)_Y$ gauge symmetry. 
The potential has accidental global symmetries in some specific cases as discussed below. 
First, let us consider the case where the $\mu$ term is absent in the potential. 
The potential then respects the global $U(1)$ symmetry which conserves the lepton number. 
Unless the lepton number is spontaneously broken the triplet field does not carry the VEV; i.e., $v_\Delta=0$. 
The terms of $\lambda_2$ and $\lambda_3$ then do not  
contribute to the four mass parameters $m_{H^{++}}$, $m_{H^+}$, $m_A$ and $m_H$, so that 
they are described by the $\lambda_4$ and $\lambda_5$ couplings. 
Therefore, two of the four are independent and the rest two are determined via Eqs.~(\ref{mass1}) and (\ref{mass2}).  
Next, we consider the case where both the $\mu$ term and the $\lambda_5$ term are zero. 
In this case, an additional global $SU(2)$ symmetry (we call this symmetry as $SU(2)_{DT}$) appears
in addition to the $U(1)$ lepton number symmetry. 
Under this $SU(2)_{DT}$ symmetry, $\Phi$ and $\Delta$ can be transformed with the different $SU(2)$ phases.  
In this case, only the $\lambda_4$ coupling can contribute to the masses of the triplet-like Higgs bosons. 
Therefore, all these masses are degenerate; i.e., $m_{H^{++}}=m_{H^+}=m_A=m_H$.

\section{Renormalization of the Higgs sector}

In this section, we define the scheme for the one-loop calculations
by setting the renormalization conditions for the parameters of the HTM 
in the similar way to the case in the two Higgs doublet model constructed in Ref.~\cite{KOSY}.   

First of all, while  at the
tree level the tadpoles $T_h$ and $T_H$ are set to be zero, at the
one-loop level they are nonzero and used to eliminate the one-point
functions of $h$ and $H$. We here prepare the tadpole counter-terms
$\delta T_h$ and $\delta T_H$ by shifting the tadpoles 
as
\begin{align}
T_h \to 0 + \delta T_h, \quad T_H \to 0 +\delta T_H. 
 \end{align}
Second, there are eight parameters in the Higgs potential:  
$m_{H^{++}}$, $m_{H^+}$, $m_H$, $m_A$, $m_h$, $\alpha$, $v$ and
$v_\Delta$. 
At the one-loop level, the counter-terms for these parameters are
obtained as 
\begin{align}
  v &\to v + \delta v, \quad 
  v_\Delta \to v_\Delta + \delta v_\Delta, \quad
  \alpha \to \alpha + \delta \alpha,\quad 
  m_i^2 \to m_i^2 + \delta m_i^2, 
\end{align}
where $i$ represents $H^{++}$, $H^+$, $A$, $H$ and $h$.
In addition to these counter-terms, those for mixing angles as well as 
the wave function renormalization factors are defined by 
\begin{align}
  H^{\pm\pm} &\to \left(1 + \frac{1}{2} \delta Z_{H^{++}}
                 \right)H^{\pm\pm},\\
  \left[\begin{array}{c}
        G^\pm \\
        H^\pm \\
        \end{array}\right]
  &\to
  \left[\begin{array}{cc}
        1 + \frac{1}{2} \delta Z_{G^+} &
         \delta \beta + \delta C_{GH} \\
       - \delta \beta + \delta C_{GH} 
         &  1 + \frac{1}{2} \delta Z_{H^+}  \\
        \end{array}\right]
  \left[\begin{array}{c}
        G^\pm \\
        H^\pm \\
        \end{array}\right], \\
 \left[\begin{array}{c}
        G^0 \\
        A \\
        \end{array}\right]
  &\to
  \left[\begin{array}{cc}
        1 + \frac{1}{2} \delta Z_{G^0} &
         \delta \beta' + \delta C_{GA} \\
       - \delta \beta' + \delta C_{GA} 
         &  1 + \frac{1}{2} \delta Z_{A}  \\
        \end{array}\right]
  \left[\begin{array}{c}
        G^0 \\
        A \\
        \end{array}\right], \\
 \left[\begin{array}{c}
        h \\
        H \\
        \end{array}\right]
  &\to
  \left[\begin{array}{cc}
        1 + \frac{1}{2} \delta Z_{h} &
         \delta \alpha + \delta C_{hH} \\
       - \delta \alpha + \delta C_{hH} 
         &  1 + \frac{1}{2} \delta Z_{H}  \\
        \end{array}\right]
  \left[\begin{array}{c}
        h \\
        H \\
        \end{array}\right]. 
\end{align}
In order to determine these 22 counter-terms 
$\delta T_h$, $\delta T_H$, $\delta v$, $\delta v_\Delta$, 
$\delta m^2_i$, $\delta Z_i$,
$\delta Z_{G^0}$,$\delta Z_{G^+}$,
$\delta \alpha$, $\delta \beta$, $\delta \beta'$,
$\delta C_{GH}$, $\delta C_{GA}$, and $\delta C_{hH}$,
we give the renormalization conditions below. 

We first discuss $\delta v$ and $\delta v_\Delta$. 
At the tree level, the VEVs $v$ and $v_\Delta$ are
determined together with the gauge coupling constants $g$ and $g'$
for SU(2)$_L$ and U(1)$_Y$ 
by the four electroweak observables $\alpha_{\rm em}$, $G_F$, $m_Z$ and $\hat{s}_W^2$ as
\begin{align}
g^2=\frac{4\pi\alpha_{\text{em}}}{\hat{s}_W^2}, \quad
g^{\prime2}=\frac{4\pi\alpha_{\text{em}}}{\hat{c}_W^2}, \quad
v^2=\frac{1}{\sqrt{2}G_F},\quad
v_\Delta^2 = \frac{\hat{s}_W^2\hat{c}_W^2}{2\pi\alpha_{\text{em}}}m_Z^2-\frac{\sqrt{2}}{4G_F}\label{vdeltree}, 
\end{align}
where $\hat{s}_W^2$ is defined as the ratio of the coefficients of
the vector part and the axial vector part in the $Z\bar{e}e$ vertex as
in Refs.~\cite{blank_hollik,ky} with $\hat{c}_W^2=1-\hat{s}_W^2$. 
We note that 
in the HTM, because the rho parameter deviates from unity at the tree level, 
the relation of $\sin^2\theta_W=1-m_W^2/m_Z^2$ does not hold. 
Thus, we have to introduce $\hat{s}_W^2$ as the independent input parameter to describe the electroweak parameters.
Consequently, the counter-terms $\delta v$ and $\delta v_\Delta$ are determined as
\begin{align}
\frac{\delta v}{v}&=\frac{1}{2}\left[\frac{\Pi_T^{WW}(0)}{m_W^2}+\delta_{VB}\right], \\
\frac{\delta v_\Delta}{v_\Delta} &=\frac{1}{2(2\hat{s}_W^2\hat{c}_W^2m_Z^2G_F-\sqrt{2}\pi\alpha_{\text{em}})}\notag\\
&\hspace{-8mm}\times\Bigg\{2\hat{s}_W^2\hat{c}_W^2m_Z^2G_F\left[
-\left.\frac{\partial}{\partial p^2}\Pi_T^{\gamma\gamma}(p^2)\right|_{p^2=0}-\frac{2\hat{s}_W}{\hat{c}_W}\frac{\Pi_T^{\gamma Z}(0)}{m_Z^2}
+\frac{\hat{c}_W^2-\hat{s}_W^2}{\hat{c}_W^2}\left(\frac{\hat{c}_W}{\hat{s}_W}\frac{\Pi_T^{\gamma Z}(m_Z^2)}{m_Z^2}-\delta_{V}'\right)
+\frac{\Pi_T^{ZZ}(m_Z^2)}{m_Z^2}
\right]\notag\\
& -\sqrt{2}\pi\alpha_{\text{em}}\left(\frac{\Pi_T^{WW}(0)}{m_W^2}
 +\delta_{VB}\right)\Bigg\}, 
\end{align}
where $\delta_{VB}$ and $\delta_{V}'$ are the vertex and box diagram corrections to the 
muon decay process and the vertex diagram corrections to the $Z\bar{e}e$ vertex, respectively, whose explicit formulae 
are given in Refs.~\cite{blank_hollik, ky},  
and $\Pi^{AB}_T(p^2)$, for ($AB=WW$, $ZZ$, $\gamma\gamma$, $\gamma Z$) are the transverse component of the 
gauge boson two-point functions. 
The counter-terms $\delta \beta$ and $\delta \beta'$ are related to
$\delta v$ and $\delta v_\Delta$ as 
\begin{align}
\delta\beta = \frac{v_\Delta}{v}\sqrt{\frac{2}{1-2v_\Delta^2/v^2}}\left(\frac{\delta v_\Delta}{v_\Delta}-\frac{\delta v}{v}\right),\quad
\delta\beta' = \frac{v_\Delta}{v}\frac{2}{(1+2v_\Delta^2/v^2)\sqrt{1-2v_\Delta^2/v^2}}\left(\frac{\delta v_\Delta}{v_\Delta}-\frac{\delta v}{v}\right). \label{tanb}
\end{align}

The tadpole counter-terms $\delta T_h$ and $\delta T_H$ are determined
so as to eliminate the one-point function for $h$ ($\Gamma_h$) and $H$ ($\Gamma_H$) at the
one-loop level: 
\begin{align}
\Gamma_h= 0,\quad \Gamma_H=0. 
\end{align}
The counter-terms for doubly-charged Higgs bosons ($\delta m_{H^{++}}^{2}$
and $\delta Z_{H^{++}}$) are determined by the on-shell conditions as
\begin{align}
  \text{Re}\Gamma_{H^{++}H^{--}}(m_{H^{++}}^2) &= 0,\label{delmH++}\\  
 \left.  \frac{\partial}{\partial p^2}
  \text{Re}\Gamma_{H^{++}H^{--}}(p^2) \right|_{p^2=m_{H^{++}}^2} &= 1, 
\end{align}
where $\Gamma_{XY}(p^2)$ represents the two-point function for
$XY$.
The four counter-terms relevant to the singly-charged scalar states
($\delta m_{H^+}^2$, $\delta Z_{H^+}$, $\delta Z_{G^+}$ and $\delta C_{GH}$)
are determined by 
\begin{align}
 \text{Re}\Gamma_{H^{+}H^{-}}(m_{H^{+}}^2) &= 0,\label{delmH+} \\
 \Gamma_{G^{+}H^{-}}(0) &= 0,\label{mix1}\\
 \left.  \frac{\partial}{\partial p^2}
  \text{Re}\Gamma_{G^{+}G^{-}}(p^2) \right|_{p^2=0} &=
\left.  \frac{\partial}{\partial p^2}
  \text{Re}\Gamma_{H^{+}H^{-}}(p^2) \right|_{p^2=m_{H^{+}}^2} 
= 1.
\end{align}
Similarly, the four counter-terms relevant to the CP-odd scalar states 
($\delta m_{A}^2$, $\delta Z_{A}$, $\delta Z_{G^0}$ and $\delta C_{GA}$)
are determined by 
\begin{align}
 \text{Re}\Gamma_{AA}(m_{A}^2) &= 0, \label{delmA} \\
 \Gamma_{GA}(0) &= 0,\label{mix2}\\ 
 \left.  \frac{\partial}{\partial p^2}
  \text{Re}\Gamma_{G^0G^0}(p^2) \right|_{p^2=0} &=
\left.  \frac{\partial}{\partial p^2}
  \text{Re}\Gamma_{AA}(p^2) \right|_{p^2=m_A^2}
= 1.
\end{align}
Finally, the six counter-terms relevant to the CP-even scalar states 
($\delta m_{h}^2$, $\delta m_{H}^2$, $\delta \alpha$,  
$\delta Z_{h}$, $\delta Z_{H}$ and $\delta C_{hH}$)
are determined by the following on-shell conditions
\begin{align}
& \text{Re}\Gamma_{hh}(m_h^2)  = \text{Re}\Gamma_{HH}(m_{H}^2) = 
 \Gamma_{hH}(m_{h}^2) = \Gamma_{hH}(m_{H}^2) = 0,\\
& \left.  \frac{\partial}{\partial p^2} 
  \text{Re}\Gamma_{hh}(p^2) \right|_{p^2=m_h^2} =
\left.  \frac{\partial}{\partial p^2}
  \text{Re}\Gamma_{HH}(p^2) \right|_{p^2=m_{H}^2}
= 1.
\end{align}
Therefore, all the 22 counter-terms are determined completely by the
above conditions, and one-loop calculations for the other
observables are now predictable.

We note that $\Gamma_{G^+G^-}(0)=0$ and $\Gamma_{G^0G^0}(0)=0$
are automatically satisfied as a consequence of the Nambu-Goldstone theorem,
when we impose the renormalization conditions of 
$\Gamma_{G^+H^-}(0)=0$ and $\Gamma_{GA}(0)=0$ in Eqs.~(\ref{mix1})
and (\ref{mix2}), respectively.
We also note that  because of the
Ward-Takahashi identities the conditions in Eqs.~(\ref{mix1})
and (\ref{mix2}) are respectively equivalent to the conditions on the mixings
between $W^\pm$ and $H^\pm$ and between $Z$ and $A$:
\begin{align}
  \Gamma_{WH}(m_{W}^2) =  \Gamma_{ZA}(m_{Z}^2) = 0,  
 \end{align}
where $\Gamma_{WH}(m_{W}^2)$ and $\Gamma_{ZA}(m_{Z}^2)$ are
defined from the $W^\pm H^\mp$ and $ZA$ two-point functions as 
\begin{align}
  \Gamma_{WH}^\mu(p^2) = -i p^\mu \Gamma_{WH}(p^2), \quad 
  \Gamma_{ZA}^\mu(p^2) = -i p^\mu \Gamma_{ZA}(p^2). 
 \end{align}

\section{Results}

We here evaluate observables at the one-loop level
by using the renormalization method defined in the previous
section.
First, we calculate the deviation from the tree-level formula among the
masses of the triplet-like Higgs bosons in Eq.~(\ref{R_def}) 
which holds approximately in the limit of $v_\Delta^2/v^2 \ll 1$.
Then, we calculate the one-loop contribution of the triplet-like Higgs
bosons to the triple Higgs boson coupling $hhh$.

At the tree level, $R$ can be expressed in terms of the quartic coupling
constants as 
\begin{eqnarray}
 R  =  \frac{\frac{1}{4} \lambda_5 v^2
  +\left(\frac{2M_\Delta^2}{v^2-2v_\Delta^2} - \lambda_5 +
    \lambda_3\right) v_\Delta^2  }
  {\frac{1}{4} \lambda_5 v^2
+ \frac{2M_\Delta^2}{v^2-2v_\Delta^2}
  v_\Delta^2 }. 
\end{eqnarray}
Under the situation of $v_\Delta^2/v^2 \ll 1$, we have 
\begin{eqnarray}
 R= 1 - 4 \left(1 - \frac{\lambda_3}{\lambda_5} \right)
 \frac{v_\Delta^2}{v^2}
  + {\cal O}\left(\frac{v_\Delta^4}{v^4}\right).
\end{eqnarray}
In the limit of $v_\Delta^2/v^2 \to 0$, $R=1$, so that the squared mass of $A$
is no more an independent parameter, which is determined by  
 $(m_A^2)_{\rm tree} = 2 m_{H^+}^2 - m_{H^{++}}^2$.

The one-loop corrected quantity of $R$ can be written as 
\begin{eqnarray}
 R^{\rm loop} = 1 + \Delta R - 4 \left(1 - \frac{\lambda_3}{\lambda_5} \right)
 \frac{v_\Delta^2}{v^2}
  + {\cal O}\left(\frac{v_\Delta^4}{v^4}\right), 
\end{eqnarray}
where $\Delta R$ is the one-loop correction to $R$ in the limit of
$v_\Delta/v \to 0$, which is evaluated as
\begin{align}
 \Delta R = \frac{m_{H^{++}}^2- m_{H^+}^2}{m_{H^+}^2 -
  (m_A^2)_{\rm pole}} -1. 
\end{align}
The squared renormalized pole mass of $A$, $ (m_A^2)_{\rm pole}$, is not the input
parameter in this limit. Therefore, we cannot adopt the on-shell renormalization
condition for $A$ given in Eq.~(\ref{delmA}) in the present case.
Instead, $(m_A^2)_{\rm pole}$ is predicted in this case by  
\begin{align}
 (m_A^2)_{\rm pole} &= (m_A^2)_{\rm tree}
   - \frac{\delta T_\Delta}{v_\Delta}
   + \delta m_A^2 - \Pi_{AA}^{\rm 1PI}[(m_A^2)_{\rm tree}]\notag\\
&\simeq(m_A^2)_{\rm tree}-
\Pi_{AA}^{\rm 1PI}[(m_A^2)_{\rm tree}]
+2\Pi_{H^+H^-}^{\rm 1PI}[m_{H^+}^2]
-\Pi_{H^{++}H^{--}}^{\rm 1PI}[m_{H^{++}}^2], \label{mApole}
 \end{align}
where $\Pi_{XY}^{\rm 1PI}$ are the 1PI diagrams for the two-point functions of $XY$, and 
$T_{h,H}^{\text{1PI}}$ are the 1PI diagrams for the one-point functions of $h$ and $H$. 
Consequently, we obtain  
\begin{eqnarray}
\Delta R =
  \frac{\Pi^{\rm 1PI}_{H^{++}H^{--}}[m_{H^{++}}^2] 
      - 2 \Pi^{\rm 1PI}_{H^{+}H^{-}}[m_{H^{+}}^2] + \Pi^{\rm 1PI}_{AA}[(m_A^2)_{\rm tree}]}
      {m_{H^{++}}^2 -m_{H^{+}}^2 }. \label{eq:delR}
\end{eqnarray}
We note that $\Delta R$ is a function of three input parameters, e.g., 
$m_{H^{++}}$, $\Delta m$ ($=m_{H^{++}}-m_{H^+}$) and $m_h$, because 
this quantity is evaluated in the limit of $v_\Delta/v\to 0$. 
We here comment on the contribution of the term which is proportional to $v_\Delta^2/v^2$ to $R$.  
As we mentioned before, the upper bound of $v_\Delta$ is about 8 GeV by the electroweak data, which means 
$v_\Delta^2/v^2\lesssim 10^{-3}$. 
Therefore, this contribution to $R$ can be neglected, and main contribution to $R$ 
should come from $\Delta R$. 
In the following numerical
analysis, we take $m_h=125$ GeV and $\alpha=0$. 

\begin{figure}[t]
\begin{center}
\includegraphics[width=70mm]{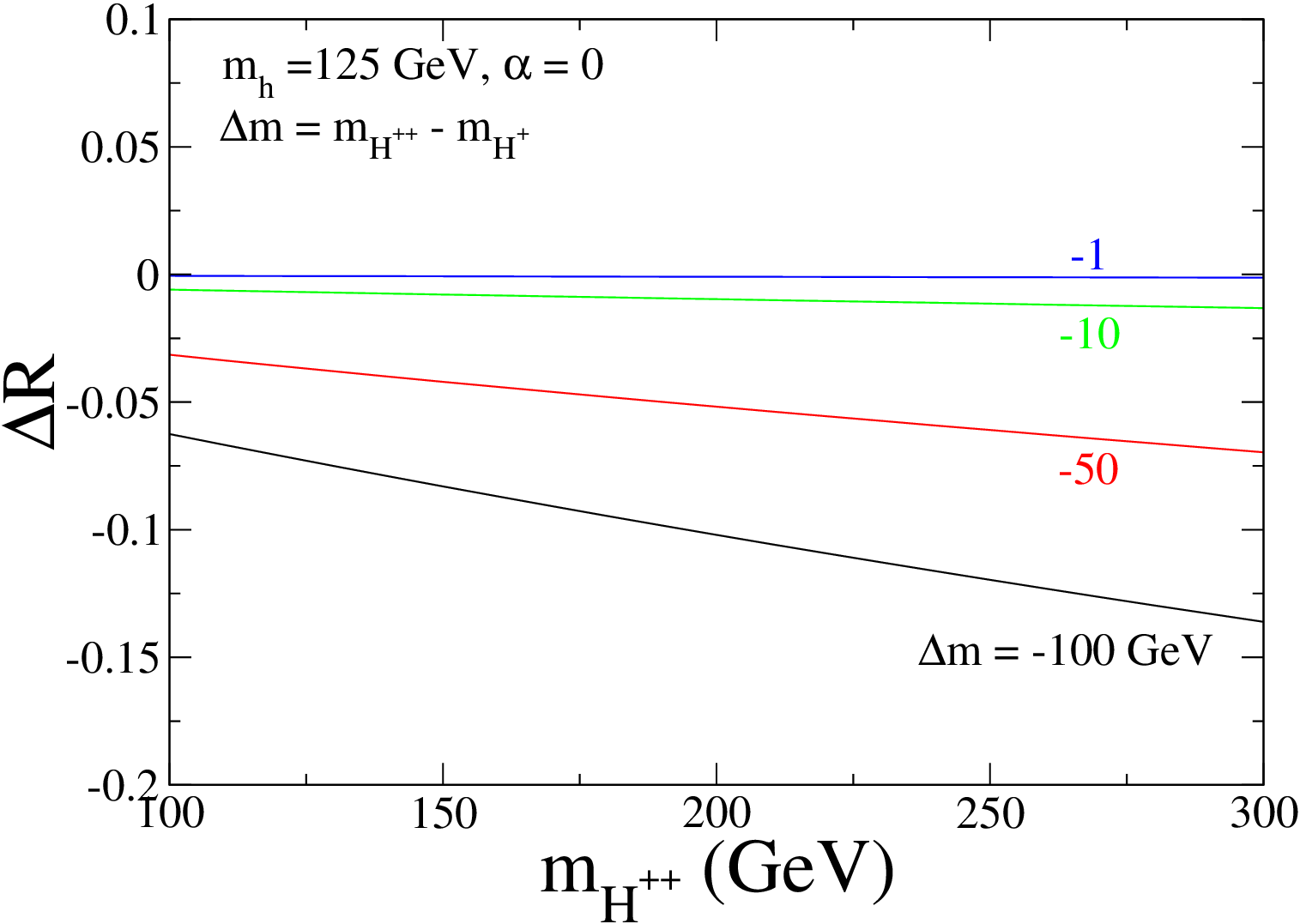}\hspace{3mm}
\includegraphics[width=70mm]{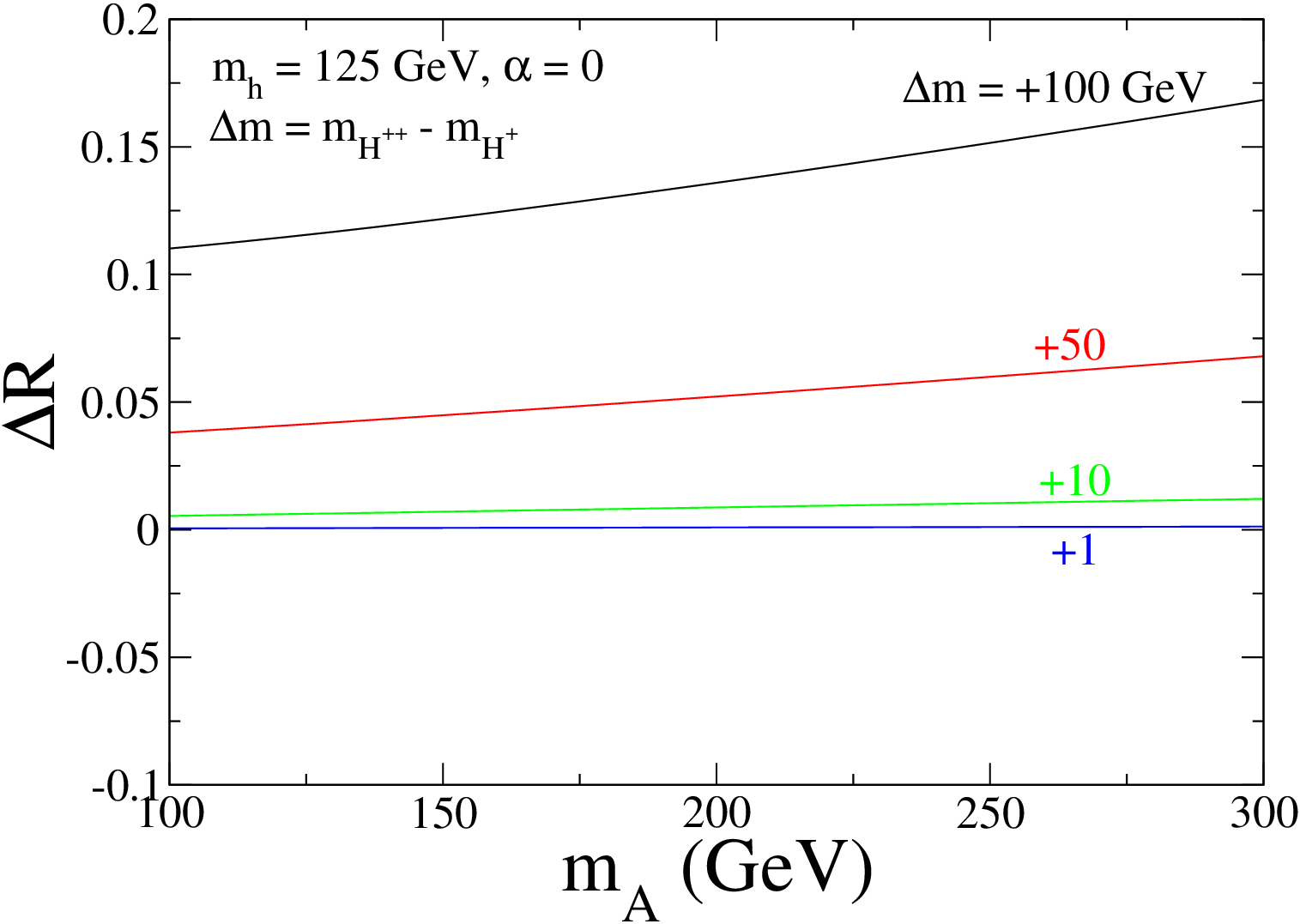}
\end{center}
\caption{
The one-loop corrections to $R$ as a function of the mass of the lightest triplet-like Higgs boson for each fixed value of $\Delta m$. 
We take $m_h=125$ GeV and $\alpha=0$, and we neglect the contributions from the terms proportional to $v_\Delta^2/v^2$.
In the left (right) figure, we show the case with the negative (positive) values of $\Delta m$. 
In the right figure, 
the horizontal axis $m_A$ is the pole mass of $A$ defined in Eq.~(\ref{mApole}). }
\label{fig:deltaR}
\end{figure}

In Fig.~\ref{fig:deltaR} (left), $\Delta R$ is shown as a function of $m_{H^{++}}$ for 
several values of $\Delta m$ ($<0$). 
From the electroweak precision data, $\Delta m = -
\mathcal{O}(100)$ GeV is favored for $m_{H^{++}} = \mathcal{O}(100)$ GeV~\cite{ky}. 
It is found that negative values of $\Delta R$ are predicted in the case with non-zero 
values of $\Delta m$. 
The magnitude of $\Delta R$ is 0.06 to 0.13 when we take $100$ GeV$<m_{H^{++}}<300$ GeV and 
$\Delta m=-100$ GeV. 
In Fig.~\ref{fig:deltaR} (right), $\Delta R$ is shown as a function 
of the pole mass of $A$ for several values of 
$\Delta m$ ($>0$) although these cases are not
necessarily favored by the electroweak precision data~\cite{ky}.
The mass of the SM-like Higgs boson $h$ is again taken to be 125 GeV. 
Positive values of $\Delta R$ are predicted.  
The magnitude of $\Delta R$ is 0.11 to 0.17
for $100$ GeV$<m_A<300$ GeV and $\Delta m=+100$ GeV.

We here give a comment regarding to the current experimental bound for the mass of the doubly-charged Higgs boson at the LHC. 
When $v_\Delta\lesssim 10^{-4}$ GeV and $\Delta m \leq 0$, the doubly-charged Higgs boson can mainly decay into the same sign dilepton. 
In such a case, the mass of the doubly-charged Higgs boson is bounded from below about 300 GeV and about 400 GeV 
by the results from the ATLAS collaboration and the CMS collaboration~\cite{mHpp}, respectively. 
Although we perform the numerical analysis in the limit of $v_\Delta/v\to 0$, 
results are not so changed even if we take $v_\Delta>10^{-4}$ GeV, because of the tiny contribution of $\mathcal{O}(v_\Delta^2/v^2)$. 
In the case of $v_\Delta>10^{-4}$ GeV and $\Delta m\leq 0$, the main decay mode of $H^{\pm\pm}$
can be the same sign diboson, so that the current bound for $m_{H^{++}}$ cannot be applied. 

Next, we discuss the one-loop contributions of the triplet-like Higgs
bosons to the triple Higgs boson coupling $hhh$ for the
SM-like Higgs boson $h$.
We calculate the $hhh$ vertex function at the one-loop level
both in the diagrammatic method in the on-shell scheme defined in the
previous section and in the effective potential method. 
The leading contribution to the deviation in the $hhh$ coupling constant $\lambda_{hhh}$ from the SM
prediction is obtained for $v_\Delta^2 \ll v^2$ as 
\begin{align}
\frac{\Delta\lambda_{hhh}}{\lambda_{hhh}^{\rm SM (loop)}} \equiv \frac{\lambda_{hhh}^{\rm HTM
 (loop)}-\lambda_{hhh}^{\rm SM (loop)}}{\lambda_{hhh}^{\rm SM (loop)}}
  \simeq  \frac{1}{12 \pi^2 v^2 m_h^2}
  ( 2 m_{H^{++}}^4 + 2 m_{H^{+}}^4 + m_A^4 + m_H^4 ). \label{eq:hhh1}
 \end{align}
Neglecting the terms proportional to $v_\Delta^2/v^2$,
$\Delta\lambda_{hhh}/\lambda_{hhh}^{\rm SM (loop)}$ is determined by three input parameters:
$m_{H^{++}}$, $\Delta m$ and $m_h$ in the same way of the calculation of $\Delta R$, 
and then Eq.~(\ref{eq:hhh1}) can be expressed as 
\begin{align}
\frac{\Delta\lambda_{hhh}}{\lambda_{hhh}^{\rm SM (loop)}}
  \simeq  \frac{1}{12 \pi^2 v^2 m_h^2}
  [ 4 m_{H^{++}}^4 + 10 (m_{H^{++}}-\Delta m)^4   -8m_{H^{++}}^2(m_{H^{++}}-\Delta m)^2]. \label{eq:hhh2}
 \end{align}
We have confirmed the correctness of this result
in both the methods of calculation; i.e., the diagrammatical method in
the on-shell scheme and the effective potential method.

We note that the one-loop contribution of the triplet-like Higgs
bosons are quartic power-like in mass, so that
it does not decouple in the large mass limit. 
The deviation from the SM prediction can be very large
when $m_h^2 \ll m_{\Delta}^2$ even under the constraint
from perturbative unitarity~\cite{unitarity_triplet},
where $m_\Delta$ represents the mass of the heaviest of all the triplet-like Higgs bosons.  
It is known that the similar effect can also appear
in the calculation of the one-loop corrected $hhh$ coupling in the two
Higgs doublet model when the masses of the scalar bosons in the loop
are given mainly by the VEV~\cite{KOSY,KKOSY}. 

\begin{figure}[t]
\begin{center}
\includegraphics[width=70mm]{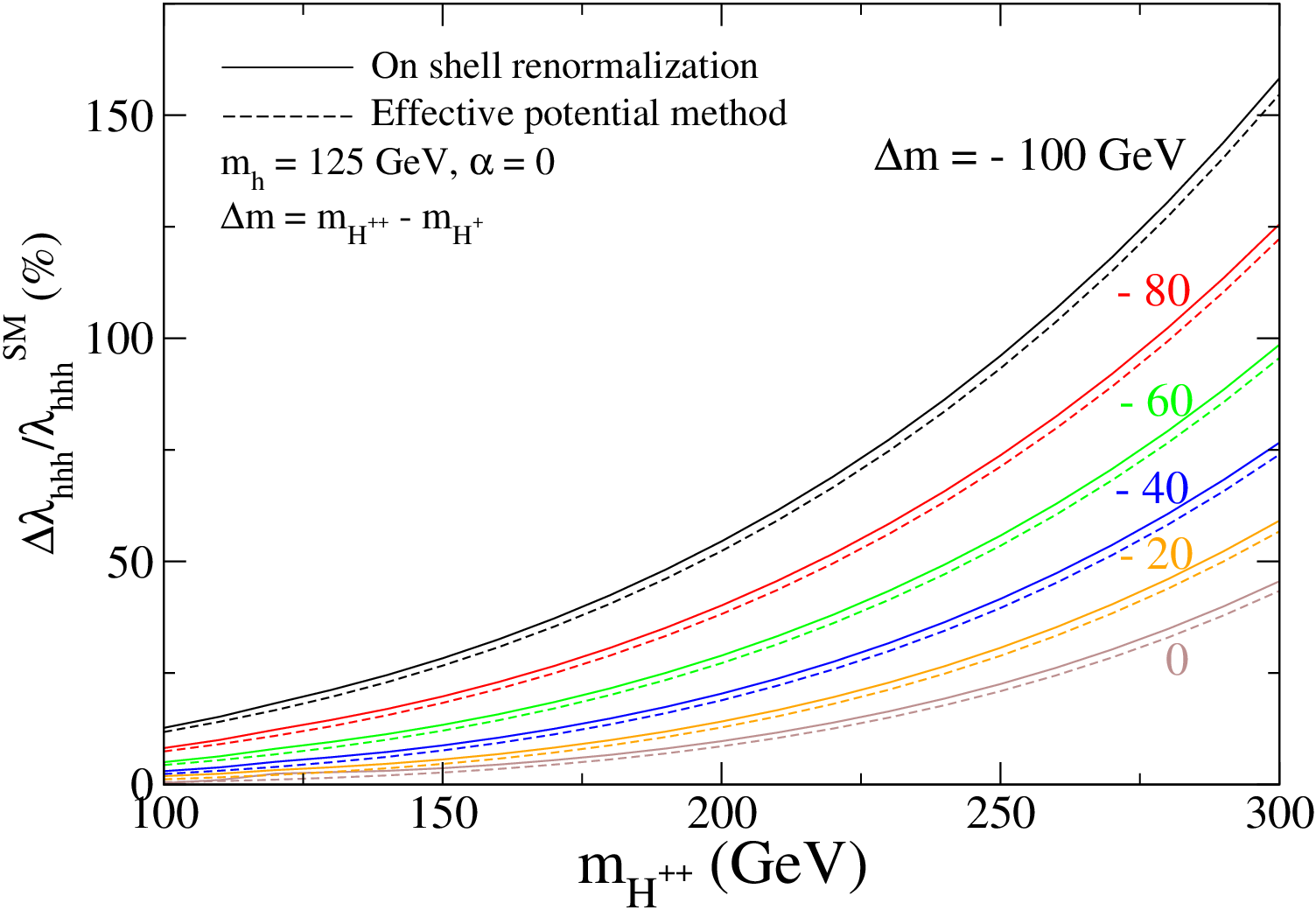} \hspace{3mm}
\includegraphics[width=70mm]{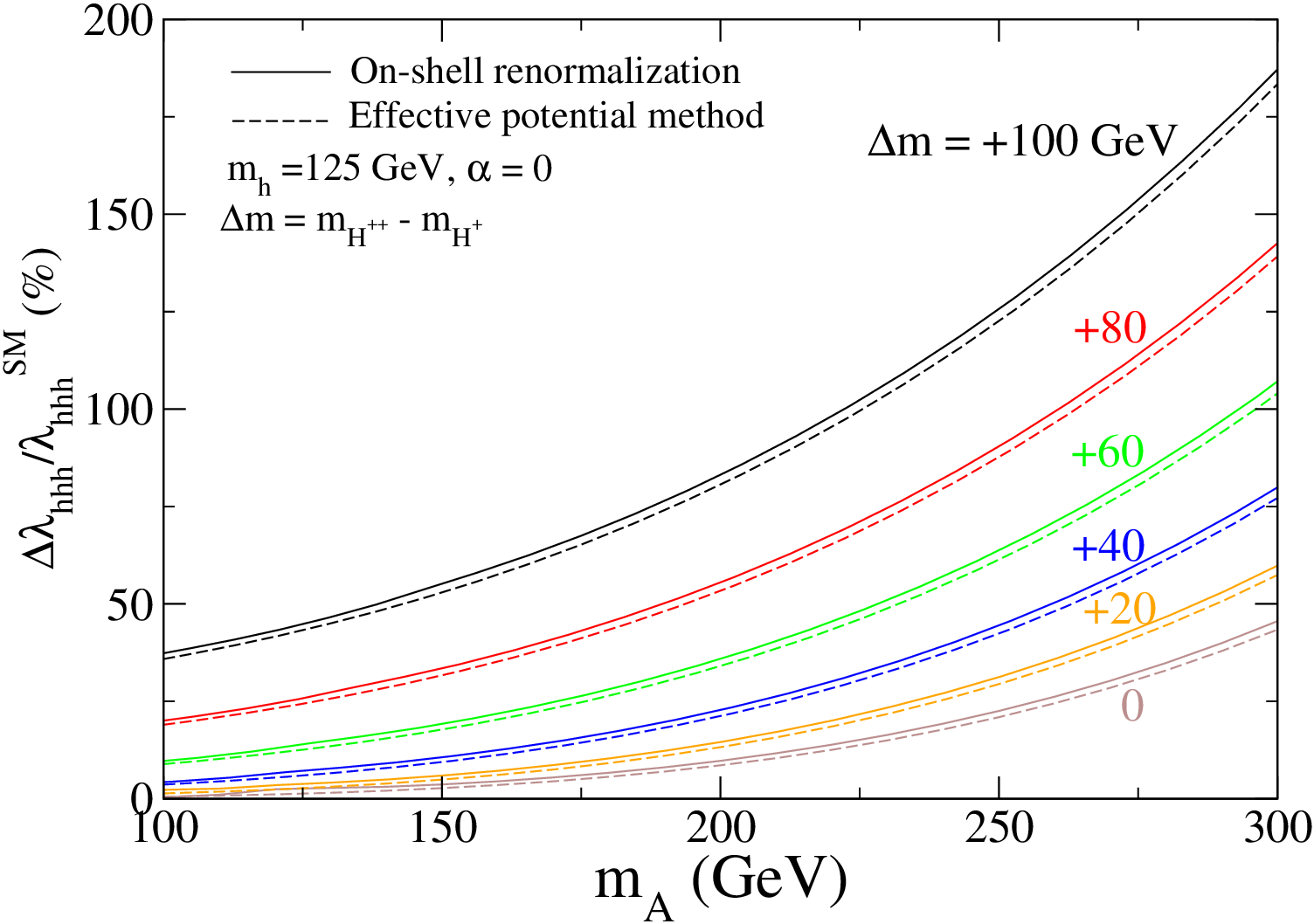}
\end{center}
\caption{The deviation of the Higgs triple coupling from the SM prediction $\Delta\lambda_{hhh}/\lambda_{hhh}^{\text{SM(loop)}}$ 
as a function of the mass of the lightest triplet-like Higgs boson for each fixed value of $\Delta m$. 
We take $m_h=125$ GeV and $\alpha=0$, and we neglect the contributions from the terms proportional to $v_\Delta^2/v^2$. 
The solid (dashed) lines represent the result by using the diagrammatical method in
the on-shell scheme (the effective potential method). 
In the left (right) figure, we show the case where $H^{++}$ ($A$) is the lightest of all the triplet-like Higgs bosons. 
In the right figure, 
the horizontal axis $m_A$ is the pole mass of $A$ defined in Eq.~(\ref{mApole}). 
}
\label{hhh}
\end{figure}

We here show the numerical results for $\Delta \lambda_{hhh}/
\lambda_{hhh}^{\rm {SM (loop)}}$ for $m_h=125$ GeV and $\alpha=0$.
In Fig.~\ref{hhh} (left), the deviation 
$\Delta\lambda_{hhh}/\lambda_{hhh}^{\text{SM(loop)}}$ is shown 
as a function of $m_{H^{++}}$ for several values of $\Delta m$ ($<0$),
where $H^{++}$ is the lightest of all the triplet-like Higgs bosons. 
The solid lines represent the result by using the diagrammatical method in
the on-shell scheme, while the dashed ones represent that by using the
effective
potential method. 
It can be seen that the results by using the two ways are almost the same. 
The small deviation comes from the effect from the wave function
renormalization.  
The deviation from the SM prediction
can be significant for large $m_{H^{++}}$ and $\Delta m$ values due to
the quartic power dependence of these parameters (see
Eq.~(\ref{eq:hhh2})).  In the case of $\Delta m=0$, the magnitude of the
deviation to be 50\% for $m_{H^{++}}=300$ GeV. 
The larger deviation is obtained for the larger value of $|\Delta m|$.  
For example when we take $\Delta m=-60$ GeV ($-100$ GeV) and
$m_{H^{++}}=300$ GeV, it amounts to about 100$\%$ (150\%).
In Fig.~\ref{hhh} (right), the similar plots are shown
as a function of $m_A$ but in the case
with positive values of $\Delta m$, where $A$ is the lightest, 
and here $m_A$ is the pole mass of $A$ defined in Eq.~(\ref{mApole}).  
The deviation from the SM prediction can also be significant.
When we take $\Delta m=+60$ GeV ($+100$ GeV) and
$m_{A}=300$ GeV, it amounts to about 100$\%$ (180\%).
We note that these large deviations satisfy the tree-level
unitarity bounds. For instance, the case with
$m_A=300$ GeV and $\Delta m=+100$ GeV implies
$m_{H^+}=413$ GeV and $m_{H^{++}}=513$ GeV.
We have confirmed that this parameter set satisfies the unitarity bounds~\cite{unitarity_triplet}.

Finally, at the ILC, the triple Higgs boson coupling would be expected to be measured
with several times 10~\% accuracy, so that the deviation in the $hhh$ coupling in
the HTM suggested by the electroweak precision data could be
detectable.


\section{conclusions}
We have investigated the one-loop renormalization for the Higgs potential in the HTM by using 
the on-shell scheme.  
We have studied how the  
characteristic relation among the masses of the triplet-like Higgs bosons 
($R\simeq 1$) 
is changed at the one-loop level. 
We have found that $R$ can be modified by the radiative corrections around $10\%$ for $v_\Delta^2/v^2 \ll 1$. 
We have also calculated the renormalized triple Higgs boson coupling in the HTM and its deviation 
from the SM prediction. 
It has been found that for the parameter sets suggested by the electroweak precision data 
the magnitude of the deviation can be as large as $\mathcal{O}(100)\%$ 
due to the quartic power-like dependence of the masses of the triplet-like Higgs bosons. 
Therefore, 
by measuring the mass spectrum of the triplet-like Higgs bosons and also 
the triple Higgs boson coupling accurately, 
the HTM can be tested at future collider experiments. 
Detailed analysis is shown elsewhere~\cite{akky_full}. 
\\\\
\noindent
$Acknowledgments$

The work of M.A. was supported in part by Grant-in-Aid for Scientific Research for Young Scientists (B), 
No. 22740137. 
The work of S.K. was supported in part by Grant-in-Aid for Scientific Research, Nos. 22244031 and 23104006. 
K.Y. was supported by Japan Society for the Promotion of Science.



\end{document}